
\newskip\oneline \oneline=1em plus.3em minus.3em
\newskip\halfline \halfline=.5em plus .15em minus.15em
\newbox\sect
\newcount\eq
\newbox\lett

\def\simlt{\mathrel{\lower2.5pt\vbox{\lineskip=0pt\baselineskip=0pt
           \hbox{$<$}\hbox{$\sim$}}}}
\def\simgt{\mathrel{\lower2.5pt\vbox{\lineskip=0pt\baselineskip=0pt
           \hbox{$>$}\hbox{$\sim$}}}}

\newdimen\short
\def\adv{\global\advance\eq by1}
\def\set#1#2{\setbox#1=\hbox{#2}}
\def\nextlet#1{\global\advance\eq by-1\setbox
                \lett=\hbox{\rlap#1\phantom{a}}}

\newcount\eqncount
\eqncount=0
\def\equn{\global\advance\eqncount by1\eqno{(\the\eqncount)} }
\def\put#1{\global\edef#1{(\the\eqncount)}           }

\def\tntf{1}
\def\tnpf{2}
\def\tnps{3}
\def\roi{5}
\def\marshak{7}
\def\tsai{10}
\def\tpc{14}
\def\suzuki{13}
\def\goun {12}
\def\lehm {6}

\def\aleph {15}
\def\pdg {11}
\def\vinh {8}
\def\tntt {4}
\def\tnpt {9}

\def\kuhn {16}
\def\isgur {17}
\def\ksrf {18}
\def\long {19}
\magnification=1200
\hsize=6.0 truein
\vsize=8.5 truein
\baselineskip 14pt
\overfullrule=0 pt

\nopagenumbers
\rightline{hep-ph/9411424}
\rightline{CPTH-A336.1194}
\rightline{November 1994}
\vskip 1.0truecm

\centerline{\bf  Application of Current Algebra in Three Pseudoscalar Meson
Decays of $\tau$ Lepton}
\vskip 1.0truecm
\centerline{{\bf L. Beldjoudi} and {\bf Tran N. Truong}}
\vskip .5truecm
\centerline{{\it Centre de Physique Th{\'e}orique}
\footnote{$^*$}{\it Laboratoire Propre du CNRS UPR A.0014}}
\centerline{\it Ecole Polytechnique, 91128 Palaiseau, France}
\vskip 2.5truecm
\centerline{\bf ABSTRACT}
\vskip .5truecm

The decays of $\tau \to 3\pi  \nu $ and $\tau \to \pi K^{*} \nu, K\rho \nu $
are calculated using the hard pion and kaon current algebra and assuming
the Axial-Vector meson dominance of the hadronic axial currents. Using the
experimental data on their masses and widths, the $\tau$ decay branching ratios
into these channels are calculated and found to be in a reasonable
agreement with the
experimental data. In particular, using the available Aleph data on the
$3\pi$ spectrum,
we determine the $A_1$ parameters, $m_A=1.24\pm 0.02 GeV$, $\Gamma
_A=0.43\pm 0.02$ GeV;
the hard current algebra calculation yields a $3\pi$  branching ratio of
$19 \pm 3 \%$.

\vskip 3.5truecm

\vfill\eject

\footline={\hss\tenrm\folio\hss}\pageno=1

\vskip 1.5 cm

Application of hard current algebra in $\tau$ decay was initiated soon after
its discovery~ [\tntf, \tnpf, \tnps ]. It was also pointed out that in a
related
calculation, using the hard current algebra technique, the cross sections
for $e^{+} e^{-} \to
4\pi^{\pm}$ can be calculated and agree with the data in the 1 GeV region
within
a factor of 4 instead of a factor of 10 using the usual soft current
algebra. This technique
was
extended to the $Ke_{4}$ decays with an implementation of the unitarity in
the 2 pion
channels~[\tntt], and was later used with success in the
resolution of the $\eta \to 3\pi$ problem~[\roi].

The technique of the hard pion current algebra consists in using the PCAC and
the Lehmann Symanzik Zimmerman reduction formula~[\lehm] by contracting out
the pseudoscalar
fields.
 One
would get the expressions for single, double or triple equal time commutator
relations (ETCR) whose Fourier transforms are the physical matrix elements
involving soft pion emissions. The remaining terms, which cannot be
calculated by current algebra
technique, involve higher order in the pseudoscalar meson momenta and
therefore do not
contribute in the hypothetical world where the pseudoscalars have zero four
momenta. In
the physical world where their four momenta do not vanish, corrections must be
taken
into account in an approximate manner by using substracted dispersion
relations; the substraction constants are given by the current algebra low
energy theorems and the substraction points are made at the scales where these
current algebra low energy theorems are valid~[\marshak].

To illustrate our point let us consider the $e^{+} e^{-} \to 4\pi$. By
contracting out the 2 S-wave pions we obtain the single and the double ETCR.
The single ETCR term is related to the physical matrix element
$\tau \to 3\pi \nu $ where the axial vector meson $A_1$ dominates; the
double ETCR is related to the physical pion form factor which is dominated by
the vector meson $\rho$. This technique enables us to incorporate the effect
of the heavy fields $\rho$ and $A_1$ which are the main features of the low
energy QCD and in a way which is consistent with chiral symmetry and
experimental data.
This method is however not unique.

Instead of this approach, it is now a fashion  to use the effective Chiral
Lagrangian which treats elegantly the low energy theorems involving the
Nambu-Goldstone bosons; quantum corrections are treated perturbatively; this
method is effectively a power series expansion in momenta and hence cannot
take  into account of the resonance effect. One is  forced to accept
the fact that Chiral Perturbation Theory, as is usually practiced, described
 only the low energy tail of the resonance which is certainly not
the gross feature of the physics involved. The remedy for
this approach is to build in the theory the heavy fields $\rho$ and $A_1$
and possibly
also a heavy $\sigma$ fields as was done by a number of authors~[\vinh,
\tnpt]. It is
suggested here that the loop corrections should be treated
unsystematically in the bubble chain approximation in order to make these
heavy fields unstable in a way which is consistent with the unitarity
requirement.
This procedure is the same as the usual way of handling the W and Z
propagators in the
standard model calculation. One then can incorporate many nice features of
the old Vector Meson Dominance models but taking into account also of the
low energy chiral
properties of the pseudoscalars. This approach will be explored in the future.

The purpose of this letter is to study the simpler processes $\tau
\to 3\pi \nu $  and $\tau \to \pi K^{*} \nu, K\rho \nu $ and leave the more
complicated process $e^{+} e^{-} \to 4\pi^{\pm}$ or $\tau \to 4\pi \nu$ for a
future publication. We do not expect to achieve here the precision of the
order of $10\%$ or better
which is usually obtained for soft pion emission processes like $Kl_2$,
$Kl_3$ and $Kl_4$ or the S
wave $\pi$N scattering lengths etc. This is so because the matrix elements
depend on
the scalar product of pion momenta to that of the current which is large in
the physical
region and considerable correction has to be made in order to reach the
chiral limit.

Using the hadronic properties (widths and masses) of the
Axial-mesons $A_1$, $Q_1$ and $Q_2$ and treat them as unstable particles in
the usual way, their
branching ratios and spectra in $\tau$ decays are calculated and found to
be in good
agreement with the experimental data.

We begin first by recalling the well-known formula [\tsai] for the ratio
$$\eqalign{R_{H}&= {\Gamma (\tau \to H^{-} \nu)\over \Gamma (\tau \to e \nu
\nu)}
\cr
&={6\pi\over m_{\tau}^2 }\pmatrix{\cos ^2\theta
_{c}\cr \sin ^2\theta _{c}\cr } \int\limits_{m_{H}^2}
^{m_{\tau}^2} dQ^2 (1-{Q^2\over m_{\tau}^2})^2\left(a_{0}(Q^2)+(1+2{Q^2\over
m_{\tau}^2})a_{1}(Q^2)\right) \cr } \eqno (1) $$
 where $a_{0}(Q^2)$ and $a_{1}(Q^2)$ are respectively the spin 0 and
spin 1 part of the hadronic spectral functions and $\theta _{c}$ is the Cabbibo
angle. The expression multiplying with $\cos ^2\theta _{c}$ is for the
$\Delta S=0$
 hadronic tau
decay and that multiplying with $\sin ^2\theta _{c}$ is for
$\Delta S=1$ hadronic tau decay.

We want to study the matrix element for the axial vector hadronic current
involving three pseudoscalar mesons. For clarity,  we first make the simplified
approximation that the system of three pseudoscalar mesons can be
represented by a
pseudoscalar and a vector meson. This assumption is reasonable because in
the usual angular momentum decomposition one pair of the pseudoscalars have
to be in the relative P state and will be shown to be dominated by the
vector meson. We assume
furthermore that the axial hadronic currents are dominated by the Axial
vector mesons
$A_1$, $Q_1$ and $Q_2$, just the same as the vector hadronic currents are
dominated by the vector
mesons $\rho$ and $K^{*}$. A more precise study of $\tau\to 3\pi \nu$ is given,
treating the vector meson $\rho$ as a resonant $2\pi$ state.

\vskip 1.0 cm
{\bf I) $\tau \to 3\pi \nu $ Decay}
\vskip 0.5 cm

Let us first consider the $\Delta S=0$ decay. As mentionned above we
approximate the $3\pi$ state by a $\pi \rho$ state. The most general
matrix element can be written as:

$$\langle\pi^{-}(k) \rho^{0}(p)\vert A^{1-i2}_{\mu} (0)\vert0
\rangle = f_1(Q^2)\epsilon_{\mu}+\epsilon .k\left( (k+p)_{\mu} f_2(Q^2)+
(k-p)_{\mu} f_3(Q^2) \right)  \eqno (2)  $$
where $Q^2=(k+p)^2$ and $\epsilon $ is the polarisation vector of $\rho$.
$f_1$,
$f_2$, and $f_3$ are complex form factors and are only functions of $Q^2$.
Current
algebra soft pion theorem, which is obtained by taking the limit $k_{\mu}
\to 0$,
gives only information on $f_1$ but not on the other 2 form factors. In an
explicit model, it was shown that they contribute little to the $\tau \to
\pi \rho \nu $. Interested readers are refered to the original article
[\roi]. (We assume here
that the decay constant of $\pi '$ is sufficiently small and hence can be
neglected). Using
the standard low energy current algebra theorem and taking the limit $ k_{\mu}
\to 0$ we have:
$$ \lim_{k_{\mu}\rightarrow 0} \langle\pi^{-}(k) \rho^{0}(p)\vert
A^{1-i2}_{\mu} (0)\vert0\rangle =-{\sqrt 2}{f_{\rho}\over f_{\pi}}\epsilon
_{\mu}(p) \eqno (3)$$
where $f_{\pi}=93MeV$, and $f_{\rho}$ is defined by the rate of $\rho \to
 e^{+} e^{-}$. Using the experimental data [\pdg] we obtain,
$f_{\rho}=0.118 GeV^2$. This
value of $f_{\rho}$ is equivalent to writing approximatively the pion form
factor as\hfill\break
$F_{\pi}(s)= m_{\rho}^2(1+\delta s/m_{\rho}^2) / \left ( m_{\rho}^2-s-i
m_{\rho}\Gamma_{\rho}(s)\right)  $. A good fit to the
experimental data is obtained with $\delta =0.2$. In fact the more general
form of
Eq(3) reads $$ \lim_{k_{\mu}\rightarrow 0} \langle\pi^{-}(k) \pi^{+}(q_1)
\pi^{-}(q_2)\vert A^{1-i2}_{\mu} (0)\vert0\rangle =-{{\sqrt 2}\over
f_{\pi}}F_{\pi}(s)(q_1-q_2) _{\mu}\eqno (4)$$ For convenience we shall first
use
Eq(3). The $3\pi$  matrix element below the $\rho \pi$ threshold can be
straightforwardly obtained from Eq(4).
 Using Eq(2) in (3) we have:
$$ f_1(m_{\rho}^2)=-{\sqrt 2}{f_{\rho}\over f_{\pi}}\eqno (5)$$
Let us start with the narrow width approximation for the $A_1$ propagator.
Using $A_1$ dominance for
the form factor we have:

$$ f_{1}(Q^2) = -{\sqrt 2}{ f_{\rho}\over f_{\pi} } { (m_{A}^2-m_{\rho}^2)
\over m_{A}^2-Q^2 }\eqno (6)$$
The generalisation of Eq[6] to take into account of the unstable nature of
$A_1$ can be
straightforwardly made. Using the $A_1$ dominance hypothesis for the axial
current, the general expression for $f_1(Q^2)$ is:
$$ f_{1}(Q^2) = -{\sqrt 2}{ f_{\rho}\over f_{\pi} } { m_{A}^2-m_{\rho}^2-\pi
(m_{\rho}^2)  \over m_{A}^2-Q^2-\pi (Q^2) } \eqno (7)$$
where we use the standard prescription for describing an unstable particle,
with
$\pi (Q^2)$ being the self energy operator of the $A_1$ resonance and is
obtained by the
bubble summation of the $\pi \rho$ intermediate states, similarly to the
treatment of the W and Z propagators in the standard model. In order to have
the usual Breit Wigner description of a resonance, we must make a twice
substracted dispersion relation with $Re[\pi(m_{A}^2)]= Re[\pi
'(m_{A}^2)]=0$ [\goun] where
$m_A$ is the $A_1$ mass:

$$\eqalign{&Re[\pi (Q^2)]= {(Q^2-m_A^2)^2\over
\pi}P\hskip -2 em \int\limits_{(m_\pi+m_{\rho})^2}^{\infty}\hskip -2 em
dz{Im[\pi
(z)]\hskip -1.1mm-\hskip -1.1mm Im[\pi(m_A^2)]\hskip -1.1mm-\hskip -1.1mm
(z-m_A^2)Im[\pi ' (m_A^2)]\over (z-m_A^2)^2(z-Q^2)} \hskip 2mm (8-a)\cr &
Im[\pi (Q^2)]={ g_{A\rho\pi}^2\over 8\pi } {{\sqrt
{\lambda (Q^2,m_{\rho}^2,m_{\pi}^2)}}\over Q^2}\left(1+{\lambda
(Q^2,m_{\rho}^2,m_{\pi
}^2) \over 12m_{\rho}^2 Q^2}\right) \hskip 2.7cm (8-b) } $$
where we define the $\pi^{0} \rho ^{+} A_1^{-}$ vertex as
$g_{A\rho\pi}\epsilon (A)
.\epsilon (\rho)$, $\lambda
\left(Q^2,m_{\rho}^2,m_{\pi}^2\right)=(Q^2-(m_{\rho}+m_{
\pi})^2)(Q^2-(m_{\rho}-m_{\pi})^2)$, and P stands for the principal part
integration. The dispersion
integral (8-a) can be written as:

$$ Re[\pi (Q^2)]=h(Q^2)-h(m_A^2)-(Q^2-m_A^2)h'(m_A^2)  \eqno (9) $$

where $$ h(Q^2) ={Q^4\over
\pi}P\int\limits_{(m_\pi+m_{\rho})^2}^{\infty}dz{Im[\pi (z)]\over
z^2(z-Q^2)}$$

$$ h(Q^2)=-{ g_{A\rho\pi}^2\over 8\pi }\left( {I_1(Q^2)\over 12m_{\rho}^2
}+{5m_{\rho}^2-m_{\pi}^2\over 6m_{\rho}^2 }I_2(Q^2)+{
(m_{\rho}^2-m_{\pi}^2)^2\over 12m_{\rho}^2
}I_3(Q^2) \right)  $$
$I_1(Q^2)$, $I_2(Q^2)$ and  $I_3(Q^2)$ are defined in the appendix. Using
Eq[7] the spectral
functions $a_{0}(Q^2)$ and $a_{1}(Q^2)$ can be straightforwardly calculated:
$$\eqalign {&a_1(Q^2)={\vert f_{1}(Q^2)\vert ^2 \over 8\pi Q^2}{ {\sqrt
{\lambda
(Q^2,m_{\rho}^2,m_{\pi}^2)}}\over Q^2}\left(1+{\lambda
(Q^2,m_{\rho}^2,m_{\pi}^2)
\over 12m_{\rho}^2 Q^2} \right)\cr
&
a_0(Q^2)= { \vert f_{1}(Q^2)\vert ^2 \over 4\pi m_{\rho}^2
}(Q^2/m_A^2-1)^2 \left({{\sqrt
{\lambda (Q^2,m_{\rho}^2,m_{\pi}^2)}}\over 2Q^2}\right)^3 }\eqno (10) $$

We want now to generalize Eq(8-b) and Eq(10) to take into account of the
unstable nature of the
$\rho$ meson, i.e we want to treat it as a resonant $2\pi$ P state. Instead
of Eq(8-b) we have
now:

$$Im[\tilde\pi(Q^2)] ={ g_{A\rho \pi}^2\over 8\pi }{1\over \pi}\hskip -2 em
\int\limits_{4m_{\pi}^2} ^{(\sqrt Q^2-m_{\pi})^2}\hskip -2 em ds
{m_{\rho}\Gamma_{\rho}(s)\over
(s-m_{\rho}^2)^2+  m_{\rho}^2\Gamma_{\rho}(s)^2 }{{\sqrt
{\lambda (Q^2,s,m_{\pi}^2)}}\over Q^2}\left(1+{\lambda (Q^2,s,m_{\pi}^2)
\over 12s
Q^2}\right) \eqno (11)$$

Instead of Eq(10) we have:
$$\eqalign{&a_{1}(Q^2)={2f_{\rho}^2\over f_{\pi}^2}{1 \over 8\pi
Q^2}{1\over \pi}\hskip -2 em
\int\limits_{4m_{\pi}^2} ^{(\sqrt Q^2-m_{\pi})^2}\hskip -2 em ds
{m_{\rho}\Gamma_{\rho}(s)\over
(s-m_{\rho}^2)^2+  m_{\rho}^2\Gamma_{\rho}(s)^2 }{ {\sqrt {\lambda
(Q^2,s,m_{\pi}^2)}}\over Q^2 }\cr &
\qquad\qquad\qquad\qquad\qquad\qquad \left(1+{\lambda
(Q^2,s,m_{\pi}^2) \over 12s Q^2}\right)\vert {m_{A}^2-s-\tilde\pi (s)\over
m_{A}^2-Q^2-\tilde\pi (Q^2) }\vert ^2\cr
&
a_0(Q^2)= {2f_{\rho}^2\over f_{\pi}^2}{ 1 \over 4\pi
}(Q^2/m_A^2-1)^2 {1\over \pi}\hskip -2 em
\int\limits_{4m_{\pi}^2} ^{(\sqrt Q^2-m_{\pi})^2}\hskip -2 em ds
{m_{\rho}\Gamma_{\rho}(s)\over
s(s-m_{\rho}^2)^2+  m_{\rho}^2\Gamma_{\rho}(s)^2 }\left({{\sqrt
{\lambda (Q^2,s,m_{\pi}^2)}}\over 2Q^2}\right)^3\cr
&
\qquad\qquad\qquad\qquad\qquad\qquad\qquad\qquad \vert {m_{A}^2-s-\tilde\pi
(s)\over m_{A}^2-Q^2-\tilde\pi (Q^2) }\vert ^2 }\eqno (12) $$
where $\Gamma_{\rho} (s)$ is the $\rho$ width; in terms of the coupling
constant
$g_{\rho \pi\pi}$ we have\hfill\break $m_{\rho}\Gamma_{\rho}(s)={ g_{\rho
\pi\pi}^2\over 48\pi }
s\left (\sqrt {1-{ 4m_{\pi}^2/ s}} \right)^3 $. From experimental data on
the $\rho$ width, we
obtain $g_{\rho\pi\pi}=6$.

$Re[\tilde\pi (Q^2)]$ is calculated as in eq[8-a] with $Im[\pi (Q^2)]$
replaced by
$Im[\tilde\pi (Q^2)]$ given in Eq[11]. $Re[\tilde\pi(Q^2)]$ is calculated
numerically.
The total branching ratio for 3$\pi$ is obtained by multiplying Eqs(10) and
(12)
by a factor of 2.

Using the narrow width approximation for the $\rho$ resonance, the $\pi\rho$
spectrum  for $m_A=1.2 GeV$ and $\Gamma _A=$0.35,  0.4 ,  0.45 GeV is shown in
Fig (1). The corresponding
R ratios are respectively 2, 1.8, 1.6 which are much larger than the
experimental
data. Because of the $\rho$ narrow width approximation, we cannot calculate
the $3\pi$
spectrum below the $\pi\rho$ threshold $s=(m_{\rho}+m_{\pi})^2$. Furthermore
the
$\pi\rho$ spectrum is found to be innacurate near to the $\pi\rho$ threshold.

We show below these disagreements with the experimental data are due to the
zero
width approximation for the $\rho$ resonance.

In Fig(2), $Re[\pi (Q^2)]$ in the $\pi\rho$ approximation,  and
 the more correct form $Re[\tilde\pi (Q^2)]$  are shown. It is seen that
$Re[\tilde\pi
(Q^2)]$ has a much smoother behavior than the approximate one $Re[\pi
(Q^2)]$  near the
$\pi\rho$ threshold. Note that at the current algebra low energy theorem limit
$s=m_{\rho}^2$, $Re[\tilde\pi (Q^2)]$ is smaller than  $Re[\pi (Q^2)]$.

In Fig 3, 4 and 5,  the $3\pi$
spectra are plotted as a function of the $3\pi$ invariant mass where, for
clarity, we normalise the theoretical prediction to the number of events.
The data are taken from the ALEPH group [\aleph] for illustration purpose.
Because the
experimental correction for acceptance has not been made, the experimental
results should be
considered as preliminary. 	In table 1 the values of $R_{3\pi}={\Gamma
(\tau \to 3\pi\nu)\over
\Gamma ( \tau \to e\nu\bar\nu)}$ are given as a function of $\Gamma _A$ and
$m_A$.

It is seen
that $R_{3\pi}$ is an increasing function of $m_A$ and decreasing function
of $\Gamma _A$.

If we assume that the acceptance
correction to the ALEPH data was negligible, our best fit to the $3\pi$
spectrum gives the following
ranges of $m_{A}$ and $\Gamma _{A}$: $m_A=1.24\pm 0.02 GeV$, $\Gamma
_A=0.43\pm 0.02$ GeV.

Using these results, the calculated branching ratio for $\tau\to 3\pi \nu$
is $19 \pm 3 \%$. The
central value corresponds to our best fit which is shown in fig(6). This
value agrees with
the ALEPH branching ratio $19.14\pm 0.48\pm0.44\%$.

The main difference between our results and the recent studies
[\kuhn,\isgur] of the $\tau \to 3\pi \nu$ is that in our approach we can
make more predictions due to the application of the low energy current algebra
theorem. That is, using the experimental mass and width of the axial vector
meson $A_1$, we
can predict the $\tau \to A_1\nu$ branching ratio. A more ambitious
approach to reduce the
$A_1$ parameters is to predict the $A_1$ width as a function of its mass,
similarly to the KSRF
relation to the $\rho$ meson[\ksrf]. This calculation will be dealt in a
forthcoming
paper[\long].

 \vskip 3.0 cm
{\bf II) $\tau \to K \rho \nu $ and $K^{*}\pi \nu$ Decays}
\vskip 0.5 cm

The study of the $\Delta S=1$ decays is more complicated due to the presence of
the two axial vector mesons $Q_1(1270)$ and $Q_2(1400)$. It is an experimental
fact that $Q_1$ is coupled strongly to the $\rho K$ and weakly to $\pi K^{*}$
while $Q_2$ couples strongly to $\pi K^{*}$ and weakly to K$\rho$. Using the
experimental data $B.R(Q_1 \to K\rho)=42\%$, $BR(Q_1 \to \pi K^{*})=16\%$,
$B.R(Q_2 \to K\rho)=3\%$ and $BR(Q_2 \to \pi K^{*})=94\% $ together with the
total width $\Gamma_{Q_1}=90 MeV$, $\Gamma_{Q_2}=170 MeV$ we obtain (in
unit of GeV):
$\vert g_{Q_1\rho K}\vert=2.6\pm 0.20$, $\vert g_{Q_2\rho K}\vert=0.62\pm
0.30$,
$\vert g_{Q_1\pi K^{*}}\vert=1.18\pm 0.20$,$\vert g_{Q_2\pi
K^{*}}\vert=3.66\pm 0.10$

Since $Q_1$ and $Q_2$ are a linear combination of the states from two different
octets, one
cannot make use of the SU(3) symmetry  to fix the signs of the coupling
constants g [\suzuki].

Similarly to Eq(3), we can
derive the low energy theorems for the matrix element $\langle
\rho^{0}(p) K^{-}(k)\vert A^{4-i5}_{\mu} (0)\vert0\rangle $ by taking the
kaon soft,
and the low energy theorem for the matrix element  $\langle \pi^{-} (k) K^{*0}
(p)\vert A^{4-i5}_{\mu} (0)\vert0\rangle $ by taking the pion soft.

In the narrow width approximation for $Q_1$ and
$Q_2$ resonances, we would expect to have the following low energy theorems:
$$\eqalign{&
f_{Q_1}{g_{Q_1\rho K}\over m_1^2-m_{\rho}^2 }+
f_{Q_2}{g_{Q_2\rho K}\over m_2^2-m_{\rho}^2 }= -{ f_{\rho}\over f_{K^+} }\cr
&
f_{Q_1}{g_{Q_1\pi K^{*}}\over m_1^2-m_{K^{*}}^2 }+
f_{Q_2}{g_{Q_2\pi K^{*}}\over m_2^2-m_{K^{*}}^2 }={ f_{K^{*}}\over f_{\pi ^{+}}
}\cr }\eqno (13)$$
where $f_{Q_1}$, $f_{Q_2}$ are similarly defined as $f_{\rho}$.

Eq(13) are strictly not correct because of the unitarity constraints due to the
complication of the coupled channel problem. A correct treatment of this
problem will be the subject of a separate publication. Because the experimental
data on $\tau \to K^{-}\rho^{0} \nu$ and $\tau \to \pi^{-} K^{*0}\nu$ are
very crude,
we make the simple approximation  that $Q_1$ couples only to
$\rho K$ and $Q_2$ to $K^{*}\pi$ and therefore neglect the mixing of these
2 channels. Solving
Eq[9]  for $f_{Q_1}$ and $f_{Q_2}$ we have, $f_{Q_1}=0.27 GeV^2$ and
$f_{Q_2}=0.39
GeV^2$.

Similarly to $f_1(s)$ defined in Eq(2), instead of Eq(7), we now have :

$$\eqalign{& f_{K\rho}(Q^2) = -{ f_{\rho}\over f_{K^{+}} } {
m_{1}^2-m_{\rho}^2-\pi_{1} (m_{\rho}^2)  \over m_{1}^2-Q^2-\pi_{1} (Q^2) }\cr
&
f_{K^{*}\pi}(Q^2) = { f_{K^{*}}\over f_{\pi^{+}} } {
m_{2}^2-m_{K^{*}}^2-\pi_{2}
(m_{K^{*}}^2)  \over m_{2}^2-Q^2-\pi_{2} (Q^2) } }\eqno (14)$$
where $\pi_{1}(Q^2)$ and $\pi_{2}(Q^2)$ satisfy a twice substracted
dispersion relation with\hfill\break  $Re[\pi_{1} (m_{1}^2)]=
Re[\pi_{1}'(m_{1}^2)]
=0$ and $Re[\pi_{2}(m_{2}^2)]= Re[\pi_{2}'(m_{2}^2)]
=0$. It should be noted that because Kaons are made of up, down and strange
quark, the chiral
(soft) Kaon limit must be taken simultaneously with the soft pion i.e we
must use  the $SU(3)_L
\times
SU(3)_R$ limit.
Because the phase space for $K \rho$ is tiny as compared with that of
$K^{*}\pi$ in $Q_1$ decays, we
must make a correction in the expression for $\pi_{1}(s)$ to take into
account of this special situation. More explicitly we have:

$$\eqalign{ Im[\pi_{1}(Q^2)]& =3/2{ g_{Q_{1}\rho K}^2\over 8\pi }{1\over
\pi}\hskip -2 em
\int\limits_{4m_{\pi}^2} ^{(\sqrt Q^2-m_K)^2}\hskip -2 em ds
{m_{\rho}\Gamma_{\rho}(s)\over (s-m_{\rho}^2)^2+
m_{\rho}^2\Gamma_{\rho}(s)^2 }{{\sqrt
{\lambda (Q^2,s,m_{K}^2)}}\over Q^2}\cr
&
\left(1+{\lambda (Q^2,s,m_{K}^2) \over 12s Q^2}\right)
+3/4{ g_{Q_{1}\pi K^{*}}^2\over 8\pi }{1\over \pi}\hskip -2
em\int\limits_{(m_\pi +m_K)^2}
^{(\sqrt Q^2-m_\pi)^2}\hskip -2 em ds  {m_{K^{*}}\Gamma_{K^{*}}(s)\over
(s-m_{K^{*}}^2)^2+
m_{K^{*}}^2\Gamma_{K^{*}}(s)^2 }\cr
&
 \left[{ {\sqrt {\lambda (Q^2,s,m_{\pi}^2)}}\over
Q^2}\left(1+{\lambda (Q^2,s,m_{\pi}^2) \over 12s Q^2}\right)+
{ ( m_{\pi}^2-m_{K}^2)^2 {\sqrt {\lambda (Q^2,s,m_{\pi}^2)}}^3 \over 6s\lambda
(s,m_{\pi}^2,m_{K}^2)) Q^4}  \right ]  }\eqno (15)  $$
It is a good approximation to make a $\delta$ function for the $K^{*}$
propagator on the right hand
side of the equation (11). Similarly we can make the $\delta$ function
approximation for the
$\rho$ and $K^{*}$ propagators.
 $$\eqalign{ Im[\pi_{2}(Q^2)]& =3/2{ g_{Q_{2}\rho K}^2\over 8\pi
}{ {\sqrt \lambda (Q^2,m_{\rho}^2,m_K^2)}\over Q^2}\left(1+{\lambda
(Q^2,m_{\rho}^2,m_K^2)  \over 12m_{\rho}^2
Q^2}\right)+\cr
&
3/4{ g_{Q_{2}K^{*}\pi}^2\over 8\pi }  {{\sqrt \lambda
(Q^2,m_{K^{*}}^2,m_{\pi}^2)}\over Q^2}\left(1+{\lambda
(Q^2,m_{K^{*}}^2,m_{\pi}^2) \over
12m_{K^{*}}^2 Q^2}\right) }$$ where  $\Gamma_{K^{*}}(s)$  is the  $K^{*}$
width. A straightforward
calculation gives:\hfill\break $m_{ K^{*}}\Gamma_{K^{*}}(s)={ g_{K^{*} \pi
K}^2\over  32\pi
}{{\sqrt{\lambda (s,m_K^2,m_{\pi}^2)}}^3/ s^2}$   . From the experimental
data on the
 $K^{*}$ width, we obtain $g_{K^{*} \pi K}=4.48$.

The self energy operators $\pi_1$ and $\pi_{2}$ can be straightforwardly
calculated using
dispersion relations. The corresponding spectral functions are easily
calculated.

$$\eqalign{&a_{1}(Q^2)={ f_{\rho}^2\over {f_K^2} }{1 \over 8\pi Q^2}{1\over
\pi}\hskip -2 em
\int\limits_{4m_{\pi}^2} ^{(\sqrt Q^2-m_K)^2}\hskip -2 em ds
{m_{\rho}\Gamma_{\rho}(s)\over (s-m_{\rho}^2)^2+
m_{\rho}^2\Gamma_{\rho}(s)^2 }{ {\sqrt {\lambda (Q^2,s,m_{\pi}^2)}}\over
Q^2}\cr
&
\qquad\qquad\qquad\qquad\qquad
\left(1+{\lambda (Q^2,s,m_{\pi}^2) \over 12s Q^2}\right)\vert
{m_{1}^2-s-\pi _{1} ^{~}(s)\over
m_{1}^2-Q^2-\pi _{1} ^{~}(Q^2) }\vert ^2\cr
&
\qquad\qquad+{ f_{K^{*}}^2\over {f_{\pi}^2} }{1 \over 8\pi Q^2}{1\over
\pi}\hskip -2 em\int\limits_{(m_\pi
+m_K)^2} ^{(\sqrt Q^2-m_\pi)^2}\hskip -2 em ds
{m_{K^{*}}\Gamma_{K^{*}}(s)\over (s-m_{K^{*}}^2)^2+
m_{K^{*}}^2\Gamma_{K^{*}}(s)^2 }\vert {m_{2}^2-s-\pi _2 ^{~}(s)\over
m_{2}^2-Q^2-\pi _2 ^{~}(Q^2)
}\vert ^2 \cr
&
\left[ { {\sqrt {\lambda (Q^2,s,m_K^2)}}\over
Q^2}\left(1+{\lambda (Q^2,s,m_{K}^2) \over 12s Q^2}\right)+
{ ( m_{\pi}^2-m_{K}^2)^2 {\sqrt {\lambda (Q^2,s,m_{K}^2)}}^3 \over 6s\lambda
(s,m_{\pi}^2,m_K^2) Q^4}\right]
 }\eqno (16)$$
and a similar expression for $a_0$.
Numerical calculation gives:
$$BR(\tau \to \rho^{0} K^{-}\nu)=0.1\%$$

$$BR(\tau \to \pi^{-} K^{0*}\nu)=0.4\%$$
These results are in good agreement with the values of TPC$/$Two-Gamma
collaboration~[\tpc]:$B\left (\tau \to K^{* 0}\pi^{-}\nu,neutrals\right)  =
0.51\pm 0.2\pm 0.13
$ and $B\left (\tau \to K^{-}\pi^{+}\pi^{-}\nu\right) =0.7 \pm 0.2$. The
$K^{-}\rho^{0}\nu$ mode is therefore consistent with zero.

\vskip 1.0 cm
{\bf  ACKNOWLEDGMENT}
\vskip 0.5 cm

We are grateful to Andr\'e Roug\'e and Henri Videau  for providing us
with  the Aleph Collaboration experimental data on $\tau \to 3\pi\nu$ decay.

\vskip 1.0 cm

\centerline {\bf {APPENDIX A}}
\vskip 0.5 cm
$$ I_1(s) = -{ s^2\over \pi}P\int \limits_{(m_\pi+m_{\rho})^2}^{\infty} {\sqrt
{\lambda(z,m_\pi^2,m_{\rho}^2)} \over z^2(z-s)}dz $$
$$ I_2(s) = -{ s^2\over \pi}P\int \limits_{(m_\pi+m_{\rho})^2}^{\infty} {\sqrt
{\lambda(z,m_\pi^2,m_{\rho}^2)} \over z^3(z-s)}dz $$
$$ I_3(s) = -{ s^2\over \pi}P\int \limits_{(m_\pi+m_{\rho})^2}^{\infty} {\sqrt
{\lambda(z,m_\pi^2,m_{\rho}^2)} \over z^4(z-s)}dz $$

These integrals are conveniently expressed  in terms of a generating function
$$ \psi(s)=-{\lambda(s,m_\pi^2,m_{\rho}^2)\over
2}P\int\limits_{(m_\pi+m_{\rho})^2}^{\infty}{dz \over \sqrt
{\lambda(z,m_\pi^2,m_{\rho}^2)}(z-s)}$$

For convenience we give the analytic continuation of this function to other
regions.

$$ \psi(s) =\cases { $$\eqalign { &\sqrt
{\lambda(s,m_\pi^2,m_{\rho})}\log\hskip -2mm  \left({{\sqrt
{s-(m_\pi+m_{\rho})^2}+\sqrt { s-(m_\pi-m_{\rho})^2}}\over 2\sqrt {m_\pi
m_{\rho}}} \right )
  } $$&\hskip -2mm if$\hskip 0.5mm s\ge (m_\pi+m_{\rho})^2$\cr
-\sqrt {\lambda(s,m_\pi^2,m_{\rho}^2)}\log\hskip -2mm  \left({{\sqrt
{-s+(m_\pi+m_{\rho})^2}+\sqrt { -s+(m_\pi-m_{\rho})^2}}\over 2\sqrt {m_\pi
m_{\rho}}}\right)&
\hskip -2mm if$\hskip 0.5mm s\le
( m_\pi-m_{\rho})^2$ \cr
\sqrt {\vert \lambda(s,m_\pi^2,m_{\rho}^2)\vert}\arctan \left (\sqrt{ {
s-(m_\pi-m_{\rho})^2}\over{-s+(m_\pi+m_{\rho})^2}} \right )&if not\cr }  $$

$$\eqalign{& I_1(s)={2\over \pi}(\psi(s)-\psi(0)-s\psi^\prime (0)) \cr
&
I_2(s)={2\over \pi s} (\psi(s)-\psi(0)-s\psi^\prime (0) -{s^2\over
2}\psi^{\prime\prime}(0))\cr
&
I_3(s) ={2\over \pi s^2}
(\psi(s)-\psi(0)-s\psi^\prime (0) -{s^2\over 2}\psi^{\prime\prime}(0)-{s^3\over
6}\psi^{\prime\prime\prime}(0)) }$$
\vskip 6cm
{\centerline {$\Gamma_A= 0.35$}}

\vbox{ \tabskip= 0pt \offinterlineskip\def\tablerule{\noalign{\hrule}}
\halign to350pt {\strut#&\vrule#\tabskip=1em plus2em& \hfil#& \vrule#&
\hfil#& \vrule# \tabskip=0pt \cr\tablerule
&&\omit\hidewidth  mass (in GeV) \hidewidth &&
\omit\hidewidth $R_{3\pi}$  \hidewidth&\cr\tablerule

&&1.15 && 1.02&\cr\tablerule
&&1.175 && 1.12&\cr\tablerule
&&1.2 &&  1.20&\cr\tablerule
&&1.25 && 1.40 &\cr\tablerule\hfil\cr}}

{\centerline {$\Gamma_A= 0.4$}}

\vbox{ \tabskip= 0pt \offinterlineskip\def\tablerule{\noalign{\hrule}}
\halign to350pt {\strut#&\vrule#\tabskip=1em plus2em& \hfil#& \vrule#&
\hfil#& \vrule# \tabskip=0pt \cr\tablerule
&&\omit\hidewidth  mass (in GeV) \hidewidth &&
\omit\hidewidth $R_{3\pi}$ \hidewidth&\cr\tablerule

&&1.15 && 0.88&\cr\tablerule
&&1.175 && 0.96&\cr\tablerule
&&1.2 &&  1.04&\cr\tablerule
&&1.26 && 1.20&\cr\tablerule\hfil\cr}}

{\centerline {$\Gamma_A= 0.45$}}
\vbox{ \tabskip= 0pt \offinterlineskip\def\tablerule{\noalign{\hrule}}
\halign to350pt {\strut#&\vrule#\tabskip=1em plus2em& \hfil#& \vrule#&
\hfil#& \vrule# \tabskip=0pt \cr\tablerule
&&\omit\hidewidth  mass (in GeV) \hidewidth &&
\omit\hidewidth $R_{3\pi}$  \hidewidth&\cr\tablerule

&&1.15 && 0.76&\cr\tablerule
&&1.175 && 0.84 &\cr\tablerule
&&1.2 &&  0.9&\cr\tablerule
&&1.25 && 1.05 &\cr\tablerule\hfil\cr}}

 \item{{\bf Table 1}}: Results for $R_{3\pi}$ defined in (1) for different
$A_1$ parameters
.\hfill\break

\vskip 1.0 cm
\centerline {\bf {REFERENCES}}
\vskip 0.5 cm

\parskip=-3pt
\item{[{\tntf}]} T.N Pham, C.Roiesnel and T.N Truong, {\it  Phy. Rev.
Lett.}{\bf 41}(1978)371
\hfill\break

\item{[{\tnpf}]} T.N Pham, C.Roiesnel and T.N Truong, {\it Phy.Lett.}{\bf
78B}(1978)623\hfill\break

\item{[{\tnps}]} T.N Pham, C.Roiesnel and T.N Truong, {\it Phy.Lett.}{\bf
80B}(1978)119
\hfill\break

\item{[{\tntt}]} T.N Truong, {\it Phy.Lett.}{\bf 99B}(1981)154 \hfill\break

\item{[{\roi}]} C.Roiesnel and T.N Truong, {\it Nucl.Phy.} {\bf B187}
(1981)293; Ecole
Polytechnique preprint A515-0982, unpublished; C.Roiesnel, Th\`ese d'Etat,
Universit\'e de Paris
Sud (Orsay) (1982). \hfill\break

\item{[{\lehm}]} H. Lehmann, K. Symanzik and W. Zimmermann {\it Nuovo
Cimento} {\bf 1}(1955)
205.   \hfill\break

\item{[{\marshak}]} R.E. Marshak, Riazuddin and C.P. Ryan, {\it Theory of
Weak Interactions
in
Particle Physics}, Wiley-Interscience (1969). \hfill\break

\item{[{\vinh}]} M.Lacombe, B.Loiseau, R. Vinh Mau, W.N.Cottingham {\it
Phy. Rev.} {\bf
D38}(1988)1491.  \hfill\break

\item{[{\tnpt}]} M. Bando, T. Kugo, K. Yamawaki, {\it Phy. Report} (1988)219
and
{\it references therein}
\hfill\break

\item{[{\tsai}]} Y.S. Tsai, {\it Phy. Rev.} {\bf D4}(1971)2821.  \hfill\break

\item{[{\pdg}]} Particle Data Group, {\it Phy. Rev.} {\bf D45}(1992).
\hfill\break

\item{[{\goun}]} G. Gounaris and J.J Sakurai, {\it Phy. Rev. Lett.} {\bf
21}(1968)244.
\hfill\break

\item{[{\suzuki}]} M. Suzuki, {\it Phy. Rev.} {\bf D47}(1993)1252.
\hfill\break

\item{[{\tpc}]} TPC/Two-Gamma Collaboration, D.A. Bauer et {\it al}.,
Report No. LBL-33037, (1993) (umpublished). \hfill\break

\item{[{\aleph}]} Aleph Collaboration {\it Z.Phy} {\bf C}59(1990)
369.\hfill\break

\item{[{\isgur}]} N. Isgur, C. Morningstar, C.Reader {\it Phy. Rev.}  {\bf
D39}(1989)1357.\hfill\break M.G. Bowler {\it Phy.Lett.}{\bf 182B}(1981)400
\hfill\break
\item{[{\kuhn}]} J.H Kuhn, A.Santamaria {\it Z.Phy} {\bf C}48(1990)
445.\hfill\break

\item{[{\ksrf}]} K.Kawarabayashi and M. Suzuki, { \it Phy. Rev. Lett.} {\bf
16}(1966)255.\hfill\break
Riazuddin and Fayyazudddin,  {\it Phy. Rev.}{\bf 147}(1966)1071.
\hfill\break

\item{[{\long}]} L. Beldjoudi, H.Ngoc Long and T.N Truong Polytechnique
preprint

 \vskip 4.1 cm
\centerline {{\bf {FIGURE CAPTIONS}}}
\vskip 0.5 cm
\item{{\bf Fig.1}} Prediction for $\tau \to 3\pi\nu$ spectrum as a function
of the $3\pi$
invariant mass in the narrow width approximation for the $\rho$ resonance
 (2$\pi$ P state =
$\rho$). dashed/solid/dot-dashed curves correspond respectively to
$\Gamma_A$=0.35,
0.4, 0.45 GeV, the $A_1$ mass is fixed to 1.2 GeV. The experimental data
are those
of Aleph group [\aleph].\hfill\break

\item{{\bf Fig.2}} Comparaison between $Re[\tilde\pi (Q^2)]$ and $Re[\pi
(Q^2)]$
 (solid/dahed
curves) as
defined in (8-a) and (11) (in units of $GeV^2$).\hfill\break

\item{{\bf Fig.3}} Calculation of the 3$\pi$ invariant mass spectrum for
$\tau \to
3\pi \nu$ decay using $\Gamma_A=0.35 GeV$. Long-dashed/
dot-dashed/short-dashed/solid curves correspond respectively to $m_A$
=1.15, 1.175,
1.2, 1.26 GeV.\hfill\break
\item{{\bf Fig.4}} Idem using $\Gamma_A$=0.4 GeV.\hfill\break
\item{{\bf Fig.5}} Idem using $\Gamma_A$=0.45 GeV. \hfill\break
\item{{\bf Fig.6}} Our best fit for the $\tau \to 3\pi \nu$ spectrum
correspending
to $m_A=1.24$ GeV, $\Gamma_A=0.43$ GeV.\hfill\break

\end